# Thermal Properties of $CO$ Diatomic Molecule in the Presence of Energy Slope Parameter (ESP)

**COLLINS EDET\* (ORCID: 0000-0001-7762-731X)**
Cross River University of Technology (CRUTECH), Faculty of Physical Sciences, Department of Physics, Cross River State, Nigeria

**UDUAKOBONG OKORIE (ORCID: 0000-0002-5660-0289)**
Akwa Ibom State University (AKSU), Faculty of Science, Department of Physics, Akwa Ibom State, Nigeria

**AKPAN IKOT (ORCID: 0000-0002-1078-262X)**
University of Port Harcourt, Faculty of Science, Department of Physics, Rivers State, Nigeria

\*Corresponding Author's Email Address: collinsokonedet@gmail.com

## ABSTRACT

*In this research article, thermal properties of energy dependent Kratzer potential (EDKP) for CO diatomic molecule is presented. The non-relativistic energy spectra earlier obtained by Ikot et al. [7] for EDKP was utilized to numerically obtain the partition function. This partition function was then used to obtain the thermal properties (such specific heat capacity, entropy, mean energy and Helmholtz free energy) of this system numerically for negative and positive values of the energy slope parameter respectively. It is observed that the energy slope parameter regulates the behavior of the system for different values. The results of this study will find direct application in molecular physics.*

## 1. INTRODUCTION

Potential models expressed as a function of energy have been explored with wave equations more than eight decades ago [1-3]. These potential models are called energy-dependent potentials. At the start, they were in existence for a particle in the external scalar and vector fields whose time independent solutions are achieved for the spin zero case. In relativistic quantum mechanics, these potentials are seen when considering particle in an electromagnetic field [1-3]. Recently, researchers have revealed rekindled interest in the study of energy-dependent Potentials (in both relativistic and non-relativistic regime) [4-6]. Ikot *et al.* [7] have recently considered the non-relativistic analysis of diatomic molecules using energy-dependent Kratzer potential (EDKP). This potential is given as follows;

$$V(r) = -2D\left( \frac{a(1+\eta E)}{r} - \frac{1}{2}\frac{a^2(1+\eta E)}{r^2} \right) \qquad (1)$$

where $D$ is the dissociation energy and $a$ is the equilibrium internuclear length, $r$ is the inter-particle distance and $\eta$ is the energy slope parameter (ESP). We point out here that when $\eta = 0$, the EDKP reduces to the Kratzer potential. Ikot *et al.* [7] pointed out that the ESP must be positive definite in order to describe a physical system. The obtained energy equation was given as follows [7];

$$E_{n\ell} = -\frac{1}{8\mu} \frac{\left(\frac{4\mu D(1+\eta E)}{\hbar^2}\right)^2 \hbar^2 a^2}{\left[n+\frac{1}{2}+\frac{1}{2}\sqrt{1+4\ell(\ell+1)+\frac{8\mu D a^2 (1+\eta E_n)}{\hbar^2}}\right]} \qquad (2)$$

In the last decade, the prediction of the thermal properties of physical systems has been a subject of interest [8, 9]. The only attempt to study the themal properties of an energy-dependent potential was presented by Lütfüolu *et al.* [10] for the energy-dependent deformed Hulthén potential energy. But none have applied the energy-dependent potential to study the thermal properties of diatomic molecule. In view of the above, the goal of the present research article is to study the thermal properties $CO$ diatomic molecule using the EDKP. The energy equation (2) presented in ref. [7] will be utilized to achieve this goal. Moreover, this study is an extension of ref. [7].

The paper is organized as follows. In section 2, the thermal properties of the EDKP is presented. In section 3, discussion of the effects of the ESP on the thermal properties of $CO$ diatomic molecule is presented. Finally, a brief concluding remark is given in section 4.

## 2. THERMAL PROPERTIES OF EDKP

In order to study the thermal properties of the EDKP, we first evaluate the partition function. This partition function (PF) can be computed by straightforward summation over all possible vibrational energy levels accessible to the system. Given the energy spectrum (2), the partition function $Z_\eta(\beta)$ of the EDKP at finite temperature $T$ is obtained with the Boltzmann factor as [10-12];

$$Z_\eta(\beta) = \sum_{n=0}^{n_{max}} e^{-\beta E_n} \qquad (3)$$

with $\beta = \frac{1}{kT}$ and with $k$ is the Boltzmann constant.

In what follows, all thermodynamic properties of the EDKP in the presence of the ESP, such as the free energy (FE), internal energy (IE), entropy, specific heat capacity (SHC), can be obtained from the partition function (3), $Z(\beta)$. These thermodynamic functions for $CO$ diatomic molecule system can be calculated from the following expressions [10];

$$F_\eta(\beta) = -\frac{1}{\beta} \ln Z_\eta(\beta),$$
$$U_\eta(\beta) = -\frac{d \ln Z_\eta(\beta)}{d\beta},$$
$$S_\eta(\beta) = \ln Z_\eta(\beta) - \beta \frac{d \ln Z_\eta(\beta)}{d\beta}, \qquad (4)$$
$$C_\eta(\beta) = \beta^2 \frac{d^2 \ln Z_\eta(\beta)}{d\beta^2}.$$

## 3. RESULTS AND DISCUSSION

In this section, we apply our results obtained in the previous sections to study the $CO$ diatomic molecules in the presence of the ESP. The $CO$ diatomic molecule is chosen because of it wide application and studies by several authors. The experimental values of molecular constants for lowest (i.e. ground) electronic state of the CO molecule are taken from literature [8]:

$D_e = 87471.43 \left( cm^{-1} \right)$, $r_e = 1.1282 \overset{o}{A}$ and $\mu = 6.860586 \, amu$. We used the following conversions; $\hbar c = 1973.269 \, eV \, \overset{0}{A}$ and $1 amu = 931.5 \times 10^6 \, eV \left( \overset{0}{A} \right)^{-1}$ for all computations [9, 11, 12].

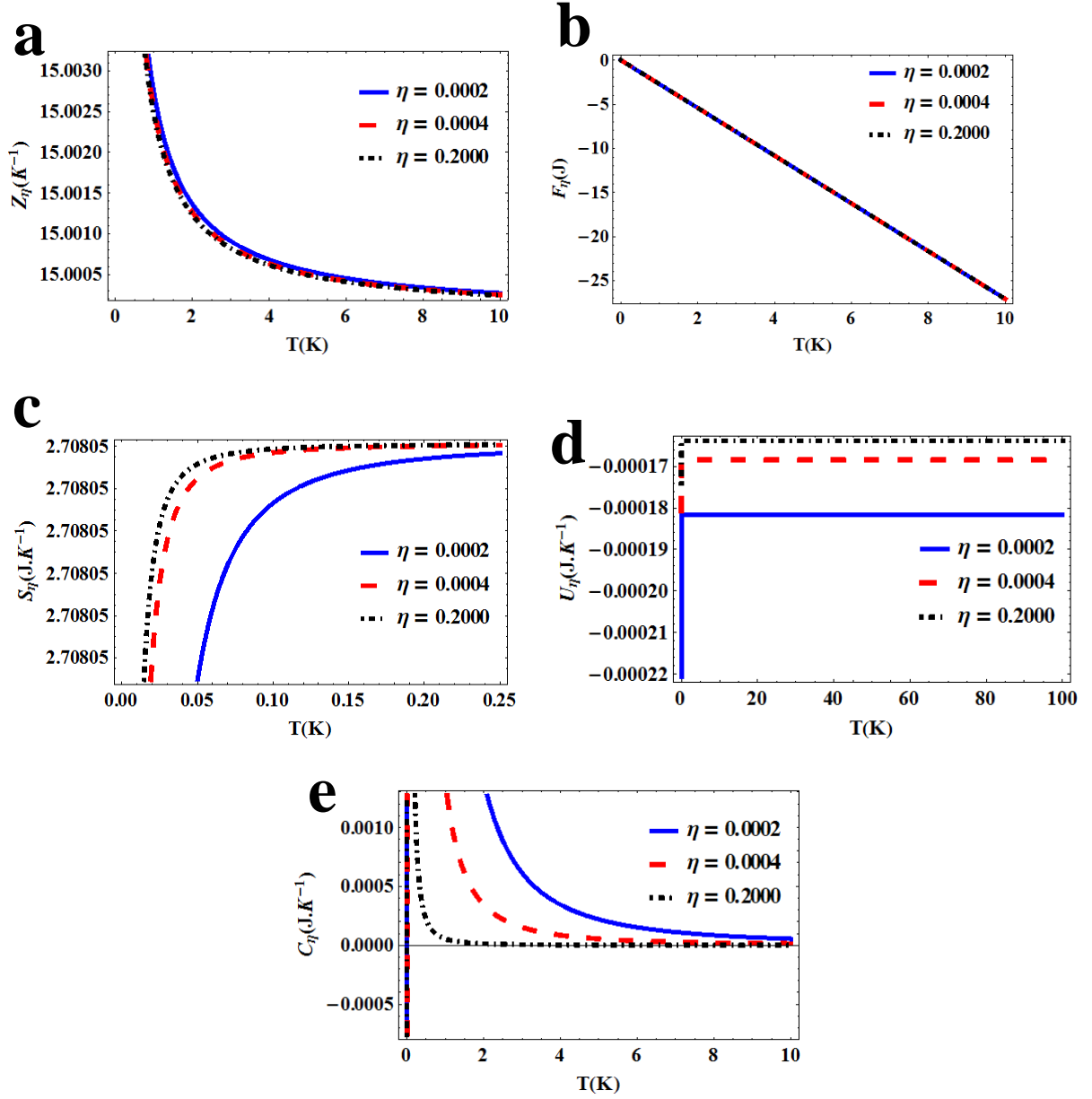

**Figure 1**: Plots of thermal properties of EDKP for $CO$ diatomic molecules in the presence of the ESP $(\eta > 0)$; **(a)** PF versus $T(K)$. **(b)** FE versus $T(K)$. **(c)** Entropy versus $T(K)$. **(d)** IE versus $T(K)$. **(e)** SHC versus $T(K)$.

In fig. 1(a), the PF decreases monotonically as $T(K)$ increases. In fig. 1(b), the FE linearly decreases with increasing $T(K)$. The entropy increases with increasing $T(K)$ as shown in fig. 1(c). The internal energy increases monotonically with increasing $T(K)$ in fig. 1(d). Fig 1(e) shows an increasing specific heat with increasing $T(K)$.

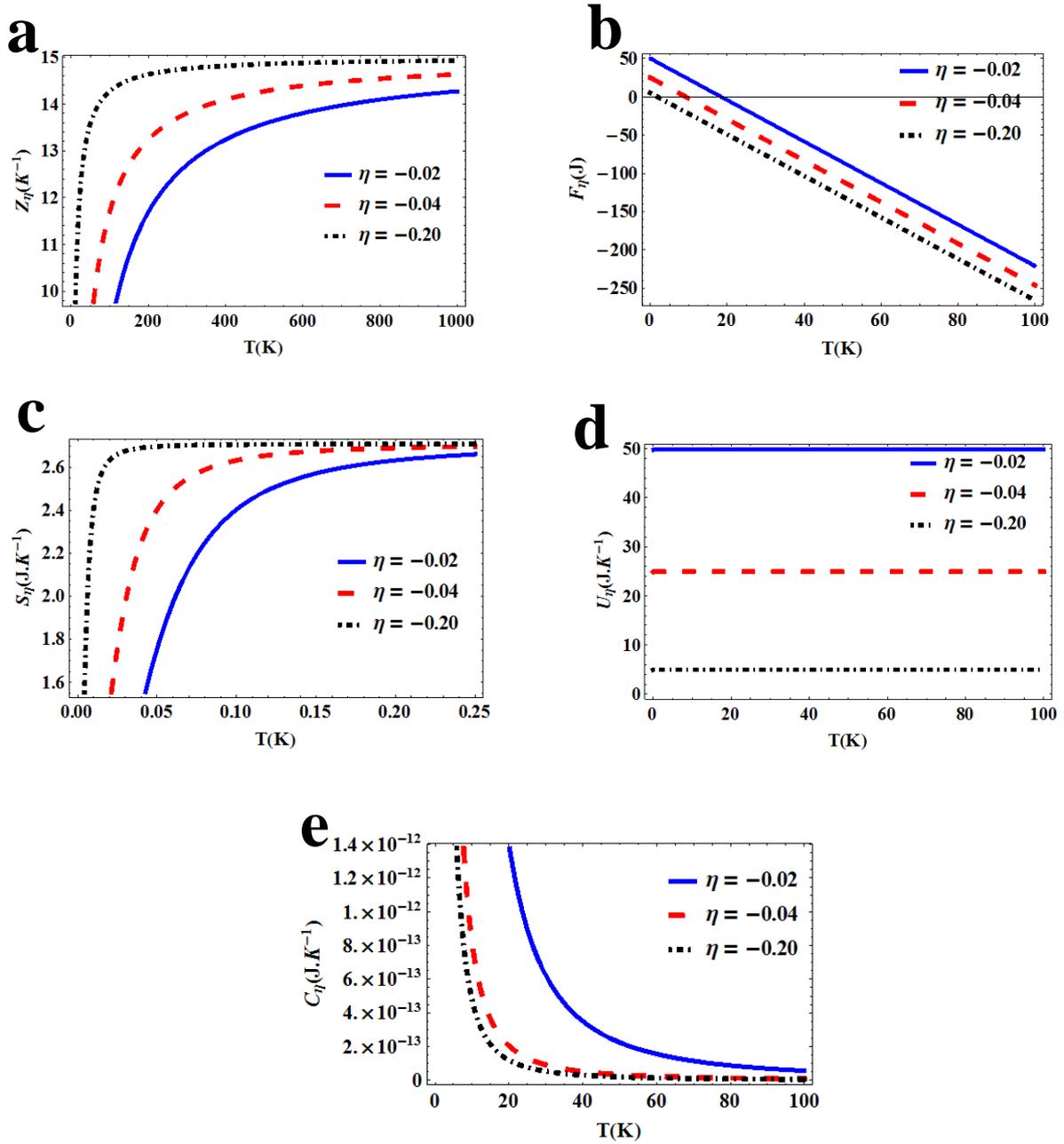

**Figure 2**: Plots of thermal properties of EDKP for $CO$ diatomic molecules in the presence of the ESP $(\eta<0)$; **(a)** PF versus $T(K)$. **(b)** FE versus $T(K)$. **(c)** Entropy versus $T(K)$. **(d)** IE versus $T(K)$. **(e)** SHC versus $T(K)$.

In fig. 2 **(a)**, the PF increases monotonically as $T(K)$ increases. In fig. 2**(b)**, the FE linearly decreases with increasing $T(K)$. The entropy increases with increasing $T(K)$ as shown in fig. 2 **(c)**. The internal energy neither increases nor decreases with increasing $T(K)$ in fig. 2 **(d)**. The IE stayed constant. Fig 2**(e)** shows an increasing specific heat with increasing $T(K)$.

Noteworthy in every curve in the SHC plots for both cases considered is an anomalous behavior (i.e. Schottky anomaly). This is because the SHC is supposed increase with increase in $T(K)$, or remains constant. However, this effect usually occurs in systems with a limited number of energy levels.

## 4. CONLUSION

In this research article, thermal properties of the energy dependent Kratzer potential (EDKP) for $CO$ diatomic molecule is studied. The goal was to scrutinize the effects of the ESP on the thermal properties this system. To achieve this goal, the non-relativistic energy spectra obtained by Ikot et al. [7] for EDKP was adopted and utilized to obtain the partition function. This was used to obtain the thermodynamic properties of the CO diatomic molecule numerically. This analysis have been carried out for $(\eta < 0)$ and $(\eta > 0)$. It is observed that the ESP regulates the behavior of the system for different values. Base on the findings of this study, it is recommend that this study should be extended to other physical systems and diatomic molecules. The study can find direct application in molecular physics.

**ACKNOWLEDGMENTS**
Collins Edet dedicates this work to his Late Father (Mr. Okon Edet Udo). In addition, Collins Edet acknowledges eJDS (ICTP).